\documentclass[12pt]{iopart}
\usepackage{iopams}  
\usepackage{graphics}
\newcommand{\vek}[1]{\mbox{\bf #1}}

\begin{document}

\title{Brane worlds}

\author{Rainer Dick}

\address{Department of Physics and Engineering Physics,
 University of Saskatchewan,\\ 116 Science Place,
 Saskatoon, SK S7N 5E2, Canada}

\begin{abstract}
Brane worlds and large extra dimensions attract
a lot of attention as possible new paradigms for spacetime.
I review the theory of gravity on 3-branes with a focus
on the codimension 1 models. However, for a new result
it is also pointed out that
the cosmological evolution of the 3-brane 
in the model of Dvali, Gabadadze and Porrati may follow
the standard Friedmann equation.
\end{abstract}


\tableofcontents

\markboth{Brane worlds}{Brane worlds}

\section{Introduction}\label{intro}

 Observation tells us that the number of macroscopic degrees
of freedom of a particle at presently accessible energies
per particle $E<1\,$TeV is three, corresponding to the
three spatial dimensions which we encounter in our
everyday lives. At the highest presently accessible energies
in collider experiments this is confirmed through
conservation of the quantum numbers of the tangential SO(3,1) symmetry,
energy and momentum: 
Momentum along a translationally symmetric brane would
be conserved anyhow in a particle scattering experiment, but 
the fact that
we do not need to account for any additional transverse momenta
in energy conservation shows that
particles do not escape into any hidden dimensions at presently
accessible energies.

Of course, the qualification with regard to the energy range also
implies the well-known fact that
small extra dimensions are well compatible with 
experimental evidence for three approximately flat macroscopic 
spatial dimensions\footnote{According to our current understanding
our knowledge of the large scale structure of the
universe is confined to the Hubble radius of order $\sim 10^{10}$
light years, if the early hot and dense phase that we see
directly in the cosmic background radiation and indirectly in the success
of the theory of primordial
nucleosynthesis emerged from an initial singularity,
or if the thermal properties of the hot and dense phase imply that any prior
information has been erased.
This qualification is understood in any statements about macroscopic
properties of spacetime.}, as has been pointed out for the first time
in the case of small {\em periodic} extra dimensions 
\cite{tk,ok}: Shifting a particle into a periodic dimension
of radius $R$ requires an energy (for $mc\ll\hbar/R$) 
\begin{equation}\label{kkeg}
\Delta E=\frac{\hbar c}{R}.
\end{equation}

Beyond Kaluza--Klein, resolving small {\em non-periodic
compact extra dimensions}
is also energetically prohibited due to the uncertainty principle,
whence a TeV-scale accelerator should be able to probe dimensions
of size\footnote{The highest energy 
single particle events that we seem to observe are the
ultrahigh energy cosmic rays with energies reaching 
almost $10^{12}\,$GeV. Our planet is hit by such a high-energetic cosmic
ray roughly once per year per $100\,\mbox{km}^2$ \cite{AGASA1,wat,AGASA2}, and
 if the observed extremely high-energetic atmospheric jets
are triggered by single particles, the propagation of these particles
through spacetime would be affected by extra dimensions of size
 $R>10^{-26}\,\mbox{cm}\approx 10^{7}\ell_{\mathrm{Planck}}$.}
 $R>10^{-17}\,\mbox{cm}\approx 10^{16}\ell_{\mathrm{Planck}}$, and
{\em non-compact finite-volume 
theories of extra dimensions}
provide phenomenologically acceptable generalizations
of the Kaluza--Klein framework
through the discreteness of the 
internal harmonic modes \cite{cw,NW,GW,OW}.

These frameworks for extra dimensions with energetically suppressed
Kaluza--Klein modes had attracted a lot of attention, 
partly for their own sake and partly for the need to include extra dimensions
in string theory, see \cite{revkk,peter,GSW,rdfp2,OW} 
and references there. Extra dimensions with TeV-scale
Kaluza--Klein modes had also been discussed in string theory
\cite{antoniadis,HW,ed,lykken,keith,zurab1}.

By the same token, the estimate $\Delta x\approx \hbar c/E$
for the resolving power seems to rule out {\em large} extra
dimensions for which the energy gap would
become so small that it should show up in particle physics experiments 
as missing energy or through excitation of a first Kaluza--Klein 
level\footnote{This remark entails
a definition of "large extra dimension": An extra dimension is {\em large}
if the corresponding Kaluza--Klein modes of matter fields
could be generated in
present day accelerators.}.

In spite of this apparent obstacle,
a framework for phenomenologically acceptable large extra dimensions
is evolving in the literature and has attracted a lot of 
attention since $\sim 1998$.
There are two main themes in this subject:
Matter must not escape into large extra dimensions,
and any viable theory of large extra dimensions
must produce a phenomenologically 
acceptable four-dimensional theory of gravity and cosmology.
  
The gravity problem is even more relevant than the escape
problem, because it is mathematically fully consistent to devise
models where matter degrees of freedom are {\it a priori} bound to
a four-dimensional
submanifold of a higher-dimensional spacetime
while the extra dimensions can only be probed by gravitons.
In the case of one large extra dimension
 the latter type of models might be described by action principles
\begin{equation}\label{scheme1}
S=\int_{\mathcal{M}_{4,1}} d^5x\,\mathcal{L}_G+
\int_{\mathcal{M}_{3,1}} d^4x\,\mathcal{L}_M,
\end{equation}
where 
the Lagrangian $\mathcal{L}_G$ would comprise all the gravity-like
degrees of freedom and $\mathcal{L}_M$ would comprise all the excitations
which can only live on the $(3+1)$-dimensional 
submanifold $\mathcal{M}_{3,1}$ (see, however, Sec.\ \ref{Action}
for the necessity to take into account extrinsic curvature terms
on $\mathcal{M}_{3,1}$ and Sec.\ \ref{dgp} for the possibility
to add the intrinsic curvature of $\mathcal{M}_{3,1}$).

 From a physical perspective, such a split of dynamical degrees of freedom
with respect to the supporting manifold may seem counter-intuitive at first
sight, but we should contemplate the possibility that nature may supply 
different degrees of freedom on different manifolds.
However, from a slightly more conservative point of view,
dynamical binding mechanisms of matter
to a four-dimensional submanifold of a higher-dimensional
spacetime have been proposed already in \cite{akama1,RuSh,mv,squires}.
In the models proposed by Akama and by Rubakov and Shaposhnikov
trapping of matter to a submanifold is implemented through the coupling
of matter to solitonic scalar fields:
Akama had used a Nielsen--Olesen vortex in $5+1$ dimensions to 
attract matter to a 3-brane \cite{akama1}, while Rubakov and
Shaposhnikov realized the matter attracting
 3-brane as a domain wall in $4+1$ dimensions, see also \cite{AH,Ru}
for recent more general discussions of solitonic binding
mechanisms. Motivated by the work on solitonic realizations,
Visser had pointed out that matter might also be 
gravitationally bound to submanifolds. Visser 
specifically proposed a model where
a $U(1)$ gauge field in $4+1$ dimensions
induces a background metric which binds
particles to a 3-brane orthogonal to the $U(1)$ gauge field \cite{mv},
and later Squires pointed out that this effect can also be due to
a bulk gravitational constant \cite{squires}.

The model of Dvali, Gabadadze and Porrati \cite{DGP} to be discussed
in Sec.\ \ref{dgp} also motivated a different approach to the
matter trapping problem, which is somewhat in between the
purely dynamical solitonic binding mechanisms and the
models where matter degrees are {\it a priori} pure brane
excitations: If there are both brane and bulk contributions
from the matter degrees of freedom to the action, then the
physics of the matter degrees of freedom
 can look four-dimensional for a certain range of parameters,
see \cite{DGS} and \cite{DR} for recent discussions of this
possibility.

The emergence of a viable four-dimensional
gravitational potential in the Newtonian limit is the primary
concern in any theory of extra dimensions with
matter degrees of freedom restricted to a
four-dimensional submanifold. The problem and its
solutions are reviewed
in Sec.\ \ref{newton}, while Sec.\ \ref{cosmology} describes
the cosmological implications of a four-dimensional world
in a higher-dimensional space probed by gravitons.

Models like (\ref{scheme1}) with $\mathcal{M}_{3,1}$ representing the
$(3+1)$-dimensional spacetime supporting matter degrees of freedom are now
widely denoted as brane worlds, and as already indicated in (\ref{scheme1}) 
the present review will focus on thin 3-branes immersed in
a $(4+1)$-dimensional spacetime, i.e.\ on models where matter degrees
of freedom
are strictly confined to a codimension 1 submanifold
$\mathcal{M}_{3,1}$, or where the energies are so low compared to a dynamical 
binding mechanism that transverse
matter excitations can be neglected.

\section{Conventions}\label{convent}

Our primary concern will be the discussion 
of dynamics in a $(4+1)$-dimensional spacetime $\mathcal{M}_{4,1}$
with matter restricted to
 a $(3+1)$-dimensional submanifold $\mathcal{M}_{3,1}$.
However, most equations and results will be expressed for 
codimension 1 hypersurfaces $\mathcal{M}_{d-1,1}$ immersed in
a spacetime $\mathcal{M}_{d,1}$.

Conventions for the metric and curvature are those of Misner, Thorne,
Wheeler \cite{MTW}, i.e.\ the metric has signature $(-+\ldots +)$, the
connection with Christoffel symbols $\Gamma^K{}_{LM}$ is always metric, 
and the Riemann and Ricci tensors are
\[
R^K{}_{LMN}=\partial_M\Gamma^K{}_{LN}
-\partial_N\Gamma^K{}_{LM}
+\Gamma^K{}_{SM}\Gamma^S{}_{LN}
-\Gamma^K{}_{SN}\Gamma^S{}_{LM}
\]
and 
\[
R_{MN}=R^K{}_{MKN},
\]
respectively.

The geodesic or proper distance from the codimension 1 
submanifold $\mathcal{M}_{d-1,1}$ will be denoted as $d^{\perp}$, and we
will define $x^{\perp}=d^{\perp}$ on one side of $\mathcal{M}_{d-1,1}$,
and $x^{\perp}=-d^{\perp}$ on the other side.
In any coordinate patch comprising a patch of $\mathcal{M}_{d-1,1}$,
$x^d\equiv x^{\perp}$ will be used as the $d$th coordinate, 
whereas the first $d-1$ coordinates $x^\mu$ cover patches 
of constant proper distance\footnote{A particular way to construct
such a Gaussian normal 
coordinate system is to first cover $\mathcal{M}_{d-1,1}$
with coordinate patches $\{x^\mu,0\le\mu\le d-1\}$ and then 
elevate these patches to a neighborhood of $\mathcal{M}_{d-1,1}$
along the perpendicular geodesics, with $x^{\perp}$ as the
$d$th coordinate. $g_{\perp\perp}=1$ and $g_{\perp\mu}=0$ 
 follow from the definition of distance and the geodesic equation.}
 from $\mathcal{M}_{d-1,1}$. This yields
Gaussian normal coordinates in a neighborhood of $\mathcal{M}_{d-1,1}$:
\begin{equation}\label{metric}
ds^2=g_{\mu\nu}dx^\mu dx^\nu+(dx^{\perp})^2,
\end{equation}
and {\it vice versa}: Due to
\[
\Gamma^{\perp}{}_{\perp\perp}=0,\,\,\Gamma^{\mu}{}_{\perp\perp}=0
\]
 $|x^{\perp}|$ in (\ref{metric})
is a geodesic distance along orthogonal trajectories to
 $\mathcal{M}_{d-1,1}$.

 The extrinsic curvature tensor in the metric (\ref{metric}) is 
\begin{equation}\label{Ktensor}
K_{\mu\nu}=-\frac{1}{2}\partial_ {\perp}g_{\mu\nu}.
\end{equation}
My convention for the $d$-dimensional Planck mass
is such that the Einstein--Hilbert action is
\[
S_{EH}=\frac{m_d^{d-1}}{2}\int d^{d+1}x\,\sqrt{-g}R.
\]
{\em Usually} this implies a $d$-dimensional Newton constant
\[
G_{N,d}=\frac{1}{2(d-1)\sqrt{\pi}^d m_d^{d-1}}
\Gamma\!\left(\frac{d}{2}\right), 
\]
i.e.\ $m_3=(8\pi G_{N,3})^{-1/2}=2.4\times 10^{18}\,$GeV
is the {\em reduced} Planck mass in $3+1$ dimensions 
(see, however, the model of Dvali, Gabadadze and Porrati 
(Sec.\ \ref{dgp}) for an exception).

With the exception of eq.\ (\ref{kkeg}) natural units
$\hbar=c=1$ are used.

\section{The Lanczos--Israel matching conditions}\label{eqs}

The unique covariant second order equation for the
metric in presence of covariantly conserved sources $\hat{T}_{MN}$
is the Einstein equation with a possible cosmological term:
\begin{equation}\label{einstein1}
R_{MN}-\frac{1}{2}g_{MN}\left(R-\frac{2\Lambda}{m_d^{d-1}}\right)
=\frac{1}{m_d^{d-1}}\hat{T}_{MN}.
\end{equation}
Spacetimes with sources confined to codimension 1 submanifolds
are no exception to this rule. 
However, with matter restricted to a codimension 1 hypersurface
\[
\hat{T}_{MN}=g_M{}^\mu g_N{}^\nu T_{\mu\nu}\delta(x^\perp)
\]
eq.\ (\ref{einstein1}) yields Einstein spaces in the bulk:
\begin{equation}\label{bulkeq}
R_{MN}=\frac{2\Lambda}{(d-1)m_d^{d-1}}g_{MN}
\end{equation}
and
a higher-dimensional version of the Lanczos--Israel matching conditions:
\begin{equation}\label{lanczoseq}
\lim_{\epsilon\to +0}\left[K_{\mu\nu}
\right]_{x^\perp =-\epsilon}^{x^\perp =\epsilon}=
\left.\frac{1}{m_d^{d-1}}\left(T_{\mu\nu}-\frac{1}{d-1}
g_{\mu\nu}g^{\alpha\beta}T_{\alpha\beta}\right)\right|_{x^\perp =0}.
\end{equation}
Here $d$ refers to the number of spatial dimensions of the embedding space,
i.e.\ $d=4$ is the case of primary interest to us. 

These matching conditions have been derived by Lanczos for the case of
singular
energy-momentum shells in general relativity
($d=3$) \cite{kl1,kl2,davis}, and a 
covariant derivation and the
geometric formulation in terms of discontinuity of extrinsic curvature 
along the singular
energy-momentum shell\footnote{According
to \cite{MTW} some of this was also anticipated by G.\ Darmois.}  
were given by Israel \cite{israel}.

 Eq.\ (\ref{lanczoseq}) implies that the geometries in the two regions
adjacent to an energy-momentum carrying codimension 1 hypersurface
differ in such a way that the extrinsic curvature of that hypersurface
is different on both sides.
Expressed in more popular terms:
What locally might be spherical from one side
might be flat from the other side. In that sense an energy-momentum carrying
codimension 1 hypersurface $\mathcal{M}_{d-1,1}$ 
could just as well be considered as a boundary
between two adjacent spacetimes $\mathcal{M}^{+}_{d,1}$ and
$\mathcal{M}^{-}_{d,1}$. $\mathcal{M}^{+}_{d,1}$ and
$\mathcal{M}^{-}_{d,1}$ are continuously connected 
along $\mathcal{M}_{d-1,1}$,
and they are smoothly connected only in those
regions of $\mathcal{M}_{d-1,1}$ where no energy-momentum currents are
present. 

Using the equations of Gauss and Codazzi, and a result of Sachs
for $R^\perp{}_{\mu\perp\nu}$ in Gaussian normal coordinates
(\cite{sachs}, cf.\ \cite{MTW}), we
can express the bulk equations (\ref{bulkeq}) in a neighborhood
of $\mathcal{M}_{d-1,1}$ in terms of intrinsic and extrinsic
curvatures of the hypersurfaces $x^\perp=$ const.\ on either side
of $\mathcal{M}_{d-1,1}$:
\begin{equation}\label{gcs1}
R_{\mu\nu}=R^{(d-1)}_{\mu\nu}+\partial_\perp K_{\mu\nu}
+2K_{\lambda\mu}K^\lambda{}_\nu
-KK_{\mu\nu}=\frac{2\Lambda}{(d-1)m_d^{d-1}}g_{\mu\nu},
\end{equation}
\begin{equation}\label{gcs2}
R_{\mu\perp}=\partial_\mu K-\nabla_\nu K^\nu{}_\mu=0,
\end{equation}
\begin{equation}\label{gcs3}
R_{\perp\perp}=g^{\mu\nu}\partial_\perp K_{\mu\nu}+K^{\mu\nu}K_{\mu\nu}
=\frac{2\Lambda}{(d-1)m_d^{d-1}}.
\end{equation}
In applications of these equations to the 
hypersurface $\mathcal{M}_{d-1,1}$ itself,
the extrinsic curvature terms should be replaced by the mean extrinsic
curvature at each point of $\mathcal{M}_{d-1,1}$:
\begin{equation}\label{meanK}
\overline{K}_{\mu\nu}=\frac{1}{2}
\lim_{\epsilon\to +0}\left[K_{\mu\nu}|_{x^\perp =-\epsilon}
+K_{\mu\nu}|_{x^\perp =\epsilon}\right].
\end{equation}

The Gauss equation (\ref{gcs1}) can be used to derive an effective
relation between the intrinsic Einstein tensor on the brane,
the local energy-momentum tensor, and the extrinsic curvatures
on the brane \cite{SMS,BCMU}.

If our codimension 1 hypersurface $\mathcal{M}_{d-1,1}$ represents an
 energy-momentum
carrying boundary with no adjacent region of spacetime on the other side
we may simply delete the corresponding extrinsic curvature term, and
(\ref{lanczoseq}) represents a boundary condition on the normal derivative 
of the metric.

\section{The action principle 
with codimension 1 hypersurfaces:
Need for the Gibbons--Hawking term}\label{Action}

Since the bulk Einstein equation (\ref{einstein1}) follows from
a bulk Einstein--Hilbert action, the natural expectation 
was that (\ref{einstein1}) with {\it a priori}
codimension 1 sources (or equivalently 
(\ref{bulkeq}) and (\ref{lanczoseq}))
could be directly derived from stationarity of
\begin{equation}\label{action1}
S_{EH}=
\int dt\int d^{d-1}\vek{x}
\int dx^\perp
\sqrt{-g}\left(\frac{m_d^{d-1}}{2}R-\Lambda\right)
\end{equation}
\[
+\left.\int dt\int d^{d-1}\vek{x}\,
 \mathcal{L}\right|_{x^\perp=0},
\]
with the brane Lagrangian $\mathcal{L}$ containing only matter
degrees of freedom and eventually intrinsic curvature terms of
the brane. 
 However, the fact that (\ref{einstein1})
follows from a bulk Einstein--Hilbert action {\it without} distinguished
submanifold does not imply that it would also follow from (\ref{action1})
{\it with} the distinguished hypersurface $\mathcal{M}_{d-1,1}$:
Every action that differs from the Einstein--Hilbert action by a complete
divergence would yield (\ref{einstein1}), but once we designate
the hypersurface $\mathcal{M}_{d-1,1}$ {\it a priori} in our action
principle, the difference in surface terms between different bulk actions
 becomes relevant, 
because the presence of energy-momentum on $\mathcal{M}_{d-1,1}$
may spoil the continuity of the surface terms across $\mathcal{M}_{d-1,1}$,
thus implying a numerical difference between the different bulk actions.
Therefore not
every bulk action which yields (\ref{einstein1}) without {\it a priori} 
designation of a hypersurface can yield 
(\ref{bulkeq}) and (\ref{lanczoseq}) from a
corresponding action principle of the sort (\ref{scheme1}), and
this applies in particular to the Einstein--Hilbert
term: Careful evaluation of the variation of $S_{EH}$
yields \cite{rdplb1,rdplb2}
\[
\delta S_{EH}=\frac{m_d^{d-1}}{2}\lim_{\epsilon\to +0}
\left(\int dt\int d^{d-1}\vek{x}
\int_{x^\perp\le -\epsilon} dx^\perp
\sqrt{-g}\delta g^{MN}\right.
\]
\begin{equation}\label{var1}
\times
\left(R_{MN}-\frac{1}{2}g_{MN}R
+\frac{\Lambda}{m_d^{d-1}}g_{MN}\right)
\end{equation}
\[
+\left.\int dt\int d^{d-1}\vek{x}
\int_{x^\perp\ge\epsilon} dx^\perp
\sqrt{-g}\delta g^{MN}\left(R_{MN}-\frac{1}{2}g_{MN}R
+\frac{\Lambda}{m_d^{d-1}}g_{MN}\right)\right)
\]
\[
+\frac{m_d^{d-1}}{2}\lim_{\epsilon\to +0}\int dt\int d^{d-1}\vek{x}
\Big[\sqrt{-g}\left(g^{MN}\delta\Gamma^\perp{}_{MN}
-g^{\perp N}
\delta\Gamma^M{}_{MN}\right)\Big]^{x^\perp=-\epsilon}_{x^\perp=\epsilon}
\]
\[
+\left.\int dt\int d^{d-1}\vek{x}\,\delta g^{MN}
\frac{\delta\mathcal{L}}{\delta g^{MN}}\right|_{x^\perp=0}
\]
\[
=\frac{m_d^{d-1}}{2}\lim_{\epsilon\to +0}\left(\int dt\int d^{d-1}\vek{x}
\int_{x^\perp\le-\epsilon} dx^\perp
\sqrt{-g}\delta g^{MN}\right.
\]
\[
\times\left(R_{MN}-\frac{1}{2}g_{MN}R
+\frac{\Lambda}{m_d^{d-1}}g_{MN}\right)
\]
\[
+\left.\int dt\int d^{d-1}\vek{x}
\int_{x^\perp\ge\epsilon} dx^\perp
\sqrt{-g}\delta g^{MN}\left(R_{MN}-\frac{1}{2}g_{MN}R
+\frac{\Lambda}{m_d^{d-1}}g_{MN}\right)\right)
\]
\[
+\frac{m_d^{d-1}}{4}\lim_{\epsilon\to +0}
\int dt\int d^{d-1}\vek{x}\Big[\sqrt{-g}
(3\delta g^{\mu\nu}\partial_\perp g_{\mu\nu}
-\delta g^{\perp\perp}g^{\mu\nu}
 \partial_\perp g_{\mu\nu}
\]
\[
+2g_{\mu\nu}\partial_\perp\delta g^{\mu\nu}
)\Big]^{x^\perp=-\epsilon}_{x^\perp=\epsilon}
+\left.\int dt\int d^{d-1}\vek{x}\,\delta g^{MN}
\frac{\delta\mathcal{L}}{\delta g^{MN}}\right|_{x^\perp=0}.
\]
The junction conditions following from $\delta S_{EH}=0$ are  
incompatible with the
junction condition (\ref{lanczoseq}). Even if we neglect the $\delta
g^{\perp\perp}$ junction term, the $\delta g^{\mu\nu}$ junction term
appears with the wrong coefficient
and a missing trace term for (\ref{lanczoseq}), and 
there appears a term proportional to 
$\partial_\perp\delta g^{\mu\nu}$
which usually has no match in $\delta\mathcal{L}/\delta g^{MN}$.
The difficulty with the Einstein action with metric discontinuities
was noticed in four dimensions in a Euclidean
ADM formalism already by Hayward and Louko \cite{HL}.

One could argue against eq.\ (\ref{var1}) that (\ref{einstein1})
implies a curvature singularity and hence
a singularity of $R$ on $\mathcal{M}_{d-1,1}$, whence additional
boundary terms should be included in (\ref{action1}) 
and (\ref{var1}). However, if
the matter on $\mathcal{M}_{d-1,1}$ is radiation dominated, then
$R$ has the same constant value 
\[
R=\frac{d+1}{d-1}\frac{2\Lambda}{m_d^{d-1}}
\]
everywhere in $\mathcal{M}_{d,1}$, and no genuine
$\delta$-function contribution
to (\ref{action1}) or (\ref{var1}) arises
classically from the Einstein--Hilbert term.
It also does not help to include an intrinsic curvature term
$\int_{\mathcal{M}_{d-1,1}}d^dx\,R^{(d-1)}$ on the brane, 
since variation of the brane intrinsic curvature scalar
cannot compensate for the $\partial_\perp\delta g^{\mu\nu}$
junction term from the bulk curvature scalar.

It was pointed out in \cite{rdplb2} that replacing the
Einstein--Hilbert term with an Einstein term 
\begin{equation}\label{LE}
\mathcal{L}_{E}=
\frac{m_d^{d-1}}{2}\sqrt{-g}
g^{MN}\left(\Gamma^K{}_{LM}\Gamma^L{}_{KN}
-\Gamma^K{}_{KL}\Gamma^L{}_{MN}\right)
\end{equation}
\[
=\mathcal{L}_{EH}-\frac{m_d^{d-1}}{2}\partial_L
\left(\sqrt{-g}g^{MN}\Gamma^L{}_{MN}-\sqrt{-g}g^{LN}\Gamma^M{}_{MN}\right),
\]
in the bulk directly yields the covariant 
equations (\ref{bulkeq},\ref{lanczoseq}).

However, a more appealing solution to the problem to derive
(\ref{bulkeq},\ref{lanczoseq}) directly from an action principle
of the kind (\ref{scheme1})
with an {\it a priori} designated hypersurface $\mathcal{M}_{d-1,1}$
employs a Gibbons--Hawking term $\sim\int_{\mathcal{M}_{d-1,1}}d^dx\,K$
\cite{GH,CR1}, see also \cite{HL,gladush,silva}.
In doing so we should take the mean extrinsic curvature $\overline{K}$
from (\ref{meanK}), because
\begin{equation}\label{Kdiff}
\lim_{\epsilon\to +0}\left[K\right]_{x^\perp =-\epsilon}^{x^\perp =\epsilon}
=
\left.-\frac{1}{(d-1)m_d^{d-1}}
g^{\mu\nu}T_{\mu\nu}\right|_{x^\perp =0}.
\end{equation}
With the proper normalization the appropriate
Gibbons--Hawking term is
\begin{equation}\label{SGH}
S_{GH}=-m_d^{d-1}\int dt\int d^{d-1}\vek{x}\,\sqrt{-g}\,
\overline{K}
\end{equation}
and variation of the metric yields
\[
\delta S_{EH}+\delta S_{GH}
=\frac{m_d^{d-1}}{2}\lim_{\epsilon\to +0}\left(\int dt\int d^{d-1}\vek{x}
\int_{x^\perp\le -\epsilon} dx^\perp
\sqrt{-g}\delta g^{MN}\right.
\]
\begin{equation}\label{var2}
\times
\left(R_{MN}-\frac{1}{2}g_{MN}R
+\frac{\Lambda}{m_d^{d-1}}g_{MN}\right)
\end{equation}
\[
+\left.\int dt\int d^{d-1}\vek{x}
\int_{x^\perp\ge\epsilon} dx^\perp
\sqrt{-g}\delta g^{MN}\left(R_{MN}-\frac{1}{2}g_{MN}R
+\frac{\Lambda}{m_d^{d-1}}g_{MN}\right)\right)
\]
\[
+\frac{m_d^{d-1}}{4}\lim_{\epsilon\to +0}\int dt\int d^{d-1}\vek{x}
\Big[
\sqrt{-g}\delta g^{\mu\nu}\left(\partial_\perp g_{\mu\nu}
-g_{\mu\nu}g^{\alpha\beta}\partial_\perp g_{\alpha\beta}\right)
\Big]^{x^\perp=-\epsilon}_{x^\perp=\epsilon}
\]
\[
+\left.\int dt\int d^{d-1}\vek{x}\,\delta g^{MN}
\frac{\delta{\mathcal L}}{\delta g^{MN}}\right|_{x^\perp=0}.
\]
$\delta S_{EH}+\delta S_{GH}=0$ 
yields exactly the Einstein condition (\ref{bulkeq}) in the
bulk and the matching condition (\ref{lanczoseq}) on the 
brane\footnote{For immersed hypersurfaces
 $S=S_{EH}+S_{GH}$ yields exactly the same variation
as the corresponding Einstein action $S_{E}$, i.e.\ the action 
without a Gibbons--Hawking term and with
an Einstein term (\ref{LE}) in the bulk. However, when the hypersurface
is a true boundary of $\mathcal{M}_{d,1}$ a further boundary term
 $\sim\sqrt{-g}\delta g^{\perp\mu}
g^{\alpha\beta}\partial_\mu g_{\alpha\beta}$ appears in $\delta S_{E}$,
which would require constant $d$-volume of the boundary.}.

\section{The Newtonian limit
on thin branes}\label{newton}

The background geometry of a spacetime satisfying 
(\ref{bulkeq},\ref{lanczoseq}) is approximately flat on 
length scales
$r\ll\left(m_d^{d-1}/|\Lambda|\right)^{1/2}$,
while on the other hand our classical calculations 
certainly become meaningless
at scales $\approx m_d^{-1}$. Hence the conditions for an ordinary flat
Newtonian limit are\footnote{E.g.\ in our spacetime with an eventual
positive cosmological constant $\Lambda<10^{-120}m_3^4$
these conditions are comfortably fulfilled on all scales where
the Newtonian limit is tested and supposed to hold \cite{will,eotw}, 
the upper limit 
being $r\ll 10^9$ light years. In this case and in our epoch
the upper limit from the bulk background matter is the same
as the limit from the maximally allowed cosmological constant.}
\begin{equation}\label{flatcon1}
\frac{1}{m_d}\ll r\ll\left(\frac{m_d^{d-1}}{|\Lambda|}\right)^{\frac{1}{2}}.
\end{equation}

On those length scales where the background geometry is approximately
flat, the gravitational potential
$U=-h_{00}/2$
of a mass distribution $\varrho(\vek{r})$ in $d\ge 3$ spatial dimensions is 
{\em usually} given by
the $d$-dimensional elliptic Green's function
for Dirichlet boundary conditions at infinity:
\begin{equation}\label{dgd}
G(\vek{r})=\frac{1}{4\sqrt{\pi}^d}
\Gamma\!\left(\frac{d-2}{2}\right)\frac{1}{r^{d-2}}
\end{equation}
through
\begin{equation}\label{nlimitd}
U(\vek{r})=-
\frac{1}{2(d-1)\sqrt{\pi}^d m_d^{d-1}}\Gamma\!\left(\frac{d}{2}\right)
\!\int d^d\vek{r}'\,\frac{\varrho(\vek{r}')}{|\vek{r}-\vek{r}'|^{d-2}}.
\end{equation}
$U(\vek{r})$ arises from the 00-component
of the $(d+1)$-dimensional Einstein equation in its linearized static form:
\begin{equation}\label{deltaU}
\Delta U(\vek{r})=\frac{1}{m_d^{d-1}}\frac{d-2}{d-1}\varrho(\vek{r}),
\end{equation}
and the corresponding potential energy of a mass $m$ is $mU(\vek{r})$.

Eqs.\ (\ref{dgd},\ref{nlimitd}) tell us that interactions 
in higher-dimensional
spacetimes are usually weaker at larger distances and stronger at shorter
distances, and Kepler's laws would not hold. {\it A priori} this sustains
in our brane models if we do not invoke special mechanisms or geometrical
constraints to ensure an $r^{3-d}$-limit for the Newton potential
on the hypersurface $\mathcal{M}_{d-1,1}$:

In models with energy-momentum bound to 
a hypersurface $\mathcal{M}_{d-1,1}$ the Newtonian
limit arises from the static weak field approximation to
(\ref{bulkeq},\ref{lanczoseq}).\\
 This yields in the bulk:
\begin{equation}\label{nlimbulk}
(\Delta+\partial_\perp^2)U(\vek{r},x^\perp)=0,
\end{equation}
and along the junction $\mathcal{M}_{d-1,1}$:
\begin{equation}\label{nlimbrane}
\lim_{\epsilon\to +0}
\left[\partial_\perp U(\vek{r},x^\perp)\right]_{x^\perp
=-\epsilon}^{x^\perp=\epsilon}
=\frac{1}{m_d^{d-1}}\frac{d-2}{d-1}\varrho(\vek{r}),
\end{equation}
or equivalently
\begin{equation}\label{nlimfull}
(\Delta+\partial_\perp^2)U(\vek{r},x^\perp)=
\frac{1}{m_d^{d-1}}\frac{d-2}{d-1}\varrho(\vek{r})\delta(x^\perp).
\end{equation}
This, of course, yields nothing but (\ref{nlimitd}) with the 
split\footnote{Things become a 
little more subtle if $\mathcal{M}_{d-1,1}$ is a compact
boundary of a spacetime with a non-periodic extra dimension. The emergence of
a $r^{-1}$-limit from a Neumann--type Green's function in such a setting is
discussed in \cite{rdplb1}.}
$\vek{r}\to\vek{r}+x^\perp\vek{e}_\perp$
and codimension 1 sources:
\begin{equation}\label{Ufull}
U(\vek{r},x^\perp)=
\frac{-1}{2(d-1)\sqrt{\pi}^d m_d^{d-1}}\Gamma\!\left(\frac{d}{2}\right)
\!\int d^{d-1}\vek{r}'\,
\frac{\varrho(\vek{r}')}{[(\vek{r}-\vek{r}')^2+{x^\perp}^2]^{(d-2)/2}},
\end{equation}
and the gravitational potential within $\mathcal{M}_{d-1,1}$ would
inherit the higher-dimensional distance law:
\begin{equation}\label{Ufull0}
U(\vek{r})=
-\frac{1}{2(d-1)\sqrt{\pi}^d m_d^{d-1}}\Gamma\!\left(\frac{d}{2}\right)
\!\int d^{d-1}\vek{r}'\,
\frac{\varrho(\vek{r}')}{|\vek{r}-\vek{r}'|^{d-2}}.
\end{equation}

However, mechanisms have been proposed in recent years to
generate a correct $(d-1)$-dimensional Newtonian limit on
$\mathcal{M}_{d-1,1}$
even for length scales $\ell^\perp$ of extra dimensions
much larger than the low-dimensional Planck
length $\ell_{d-1}=1/m_{d-1}$ or the length scales $\hbar c/E$
which ordinarily should be ruled out through accelerators:

\subsection{The observation of Arkani-Hamed, Dimopoulos and
  Dvali}\label{ADDmech}

If the extra dimension has a finite extension $\ell^\perp$ which
is well below the minimal currently accessible length 
scale $\approx 0.2\,$mm
for tests of Einstein gravity \cite{eotw}, then the
Dirichlet Green's function (\ref{dgd}) with vanishing boundary
condition for $x^\perp\to\infty$ is certainly not appropriate,
and we should expect that the
low-dimensional Newtonian potential at scales $r>\ell^\perp$
should result from a gravitational field which is quenched
over the transverse dimension \cite{ADD1,AADD,ADD2}. 
This should yield the expected
$(d-1)$-dimensional gravitational potential on $\mathcal{M}_{d-1,1}$.
 I will denote this as an ADD type mechanism. 
In that case even the fundamental quantum gravity scale
$m_4$ of the theory can be much smaller than our
4-dimensional Planck mass $m_3$ \cite{ADD1,ADD2}.

 Following \cite{ADD2} we can derive the relation between the
Newton constants in $d-1$ and $d$ spatial dimensions under the
assumption that the extra dimension has finite length $\ell^\perp$
by calculating the flux of the gravitational field
of a mass $M$ through a $d$-dimensional
cylinder of radius $r$ and length $\ell^\perp$:

 From (\ref{deltaU}) we get for $r>\ell^\perp$
\[
\frac{2\sqrt{\pi}^{d-1}}{\Gamma\!\left(\frac{d-1}{2}\right)}
\ell^\perp r^{d-2}\frac{dU}{dr}=\frac{d-2}{d-1}\frac{M}{m_d^{d-1}},
\]
i.e.\
\[
U(r)=-\frac{d-2}{2(d-1)(d-3)\sqrt{\pi}^{d-1}m_d^{d-1}\ell^\perp}
\Gamma\!\left(\frac{d-1}{2}\right)\frac{M}{r^{d-3}}.
\]
Comparison with (\ref{nlimitd}) for $d-1$ spatial dimensions
yields 
\begin{equation}\label{relADD}
(d-2)^2m_{d-1}^{d-2}=(d-1)(d-3)m_d^{d-1}\ell^\perp.
\end{equation}
 
 For $d=4$ and $\ell^\perp< 0.2\,$mm this yields a lower
bound on a 5-dimensional Planck scale which is well below $m_3$:
\begin{equation}\label{m4bound}
m_4=\left(\frac{4m_3^2}{3\ell^\perp}\right)^{\frac{1}{3}}
>2\times 10^8\,\mbox{GeV}.
\end{equation}
The result for general number $\nu$ of extra spacelike
dimensions of length $\ell^\perp$ is
\[
m_{3+\nu}=\left(
\frac{2(\nu+1)m_3^2}{(\nu+2){\ell^\perp}^\nu}\right)^{\frac{1}{\nu+2}}
>10^{(37-12\nu)/(\nu+2)}\,\mbox{GeV}.
\]

Two concentric 3-spheres of radii $a<b$
provide a model system 
where the realization of the ADD mechanism for $a,b\gg b-a$
and the realization of the higher-dimensional singularity
for $b\gg a$ can be studied analytically. This model realizes the
3-brane as a boundary of a spatial four-dimensional bulk, and therefore
the potential arises from the Green's function for Neumann
boundary conditions, but it can be written down exactly in terms
of the four-dimensional multipole expansion with azimuthal symmetry
\cite{rdplb1}.

\subsection{The Randall--Sundrum model}\label{RSmodel}

Another possibility arises if our low-dimensional
Newtonian limit is not flat, 
because a bulk cosmological term induces a
transverse length scale smaller than the length
scales tested in experimental gravity. Again 
the flat background approximation in our
calculation of the Newtonian limit would be
invalidated, and the classical approximation might
be invalidated as well. In consideration of \cite{RS1,RS2}
I will denote this as an RS type mechanism.  
Earlier discussions of the emergence
of metrics of the form
 $ds^2=\phi(x^\perp)\eta_{\mu\nu}dx^\mu dx^\nu+d{x^\perp}^2$ 
in 5-dimensional models can be found in
\cite{RuSh,merab}. $\phi(x^\perp)$ is usually denoted
as a warp factor in these models.

Randall and Sundrum have proposed 
a 3-brane with a brane tension $\lambda_3$
in a 5-dimensional bulk with a cosmological
constant $\Lambda$.
The metric
\begin{equation}\label{RSm}
ds^2=
\exp\!\left(-\frac{\lambda_3}{3m_4^3}|x^\perp|\right)
\eta_{\mu\nu}dx^\mu dx^\nu
+d{x^\perp}^2
\end{equation}
solves (\ref{einstein1}) for $d=4$ with
a bulk cosmological constant
\begin{equation}\label{RSL}
\Lambda=-\frac{\lambda_3^2}{6m_4^3}
\end{equation}
and
\[
\hat{T}_{MN}=-g_M{}^\mu g_N{}^\nu\eta_{\mu\nu}
\lambda_3\delta(x^\perp).
\]

The original setup of Randall and Sundrum consisted of a
$\mathbb{Z}_2$-symmetric configuration of two 3-branes
 embedded in $\mathcal{M}_{3,1}\times S_1$. 
If the transverse
extension $2\pi R_\circ$ is very small compared to $0.2\,$mm, 
then the argument
of ADD applies to ensure an ordinary Newtonian limit
at distances $r\gg R_\circ$. However, it was argued in
\cite{RS2} that a 3-brane in an infinitely
extended bulk (\ref{RSm})
yields a viable approximation to ordinary Newtonian gravity
in four dimensions through a trapped massless 
graviton mode (see also
\cite{kaloper,CEHS,ADDK,GRS1,LMW,CGK,LY,CH2}
 for corresponding setups of more branes and the
stabilization problem in theories with several branes).
This proposal implies
that curvature should play an important role
in the low-energy limit, and this has to be taken
into account in the discussion of graviton evolution
equations in this type of models.
The problem of the weak field expansion
around the corresponding
 curved background has been studied by many groups
\cite{SMS,GT,GKR,grojean,MVV,AIMVV,GRS1,CEH,GRS2,
CH1,zurab2,CED,GR2,NS,neupane}.

Therefore the primary problem to address in the present setting is: 
Can the transverse
distance $|x^\perp|$ from a flat 3-brane in the metric (\ref{RSm})
 exceed
$0.2\,$mm without contradicting experimental tests of Einstein
gravity, due to the curvature of the background geometry?

\setcounter{footnote}{1}
Motion along the brane in the
stationary weak field expansion around the
background geometry (\ref{RSm})
can be described by a gravitational potential\footnote{
The gravitational potential cannot fully
account for motion
in the transverse direction, if particles could leave the
brane.} (with $h_{MN}=\delta g_{MN}$)
\[
U=-\frac{1}{2}\exp\!\left(\frac{\lambda_3}{3m_4^3}|x^\perp|\right)
 h_{00}.
\]
However, as indicated above, a peculiarity arises in the discussion
of the phenomenological suitability
of the Randall--Sundrum model with a large
extra dimension: Curvature must play an essential
role if the model as proposed with a large extra dimension
is to reproduce a four-dimensional gravitational potential
at those length scales where we observe the potential.
This implies that in the calculation of $U$
 we have to expand our fields around
the curved background (\ref{RSm}) rather than around an
approximately flat section of that background.
However, gravitons are tensor particles which do not satisfy
decoupled evolution equations in a curved background,
 and it is also not possible to derive
a decoupled equation for the gravitational potential.
This is obvious also from the analogy between the problem to derive
a graviton evolution equation in a curved background and the
theory of cosmological perturbations, 
and another direct way to see non-separability 
in a curved background is to recall the
formula 
\[
\delta R_{MN}=\nabla_K\delta\Gamma^K{}_{MN}
-\nabla_N\delta\Gamma^K{}_{KM}
\]
for the first order variation of the Ricci tensor
under first order changes of the metric\footnote{Even
in a flat background Gaussian
normal coordinates would not be the best choice when it
comes to separation of the Einstein equation in the
weak field approximation.
However, as emphasized above, the problem at hand also has
a gauge independent origin in the background
curvature. A harmonic gauge for the
longitudinal coordinates $x^\mu$ (with $h^\alpha{}_\mu
=\eta^{\alpha\beta}h_{\beta\mu}$, etc.):
\[
\partial_\alpha h^\alpha{}_\mu
-\frac{1}{2}\partial_\mu h^\alpha{}_\alpha
=\frac{1}{2}\exp\!\left(-\frac{\lambda_3}{3m_4^3}|x^\perp|\right)
\partial_\mu h_{\perp\perp}
-\partial_\perp\left[
\exp\!\left(-\frac{\lambda_3}{3m_4^3}|x^\perp|\right) h_{\perp\mu}\right]
\]
is useful in separating the 2nd order
derivative terms, but the evolution of $h_{\mu\nu}$
will not decouple from
 $h_{\perp\perp}$, $h_{\perp\mu}$
and $h^\alpha{}_\alpha$. In principle, one could eliminate the latter
by solving the coupled set of equations involving
 $R_{\perp\perp}$, $R_{\perp\mu}$
and $R^\alpha{}_\alpha$, but that means trading the couplings
for non-local source terms.}

To get a qualitative understanding of the expected behavior
of the gravitational potential, we may consider
the following equation for the gravitational potential,
which arises from the diagonal terms
of $R_{00}$ for time-independent $h_{00}$:
\begin{equation}\label{RSU}
\Delta U(\vek{r},x^\perp)+
\exp\!\left(\frac{\lambda_3}{3m_4^3}|x^\perp|\right)\partial_\perp
\left[\exp\!\left(-\frac{2\lambda_3}{3m_4^3}|x^\perp|\right)
\partial_\perp U(\vek{r},x^\perp)\right]
\end{equation}
\[
=\frac{2}{3m_4^3}\varrho(\vek{r})\delta(x^\perp).
\]
Note that the differential operator on the left hand side is just
$\exp(-\lambda_3|x^\perp|/(3m_4^3))$ times the scalar covariant Laplacian
in the metric (\ref{RSm}), applied to a time-independent field.

Inspecting this equation for an extended, weakly $\vek{r}$-dependent
source shows that the potential orthogonal to the source
evolves exponentially
\[
U(x^\perp)\sim \frac{1}{\lambda_3}
\exp\!\left(\frac{2\lambda_3}{3m_4^3}|x^\perp|\right).
\]
This indicates that
the resulting gravitational potential of a mass distribution
$\varrho(\vek{r})$ on the negative tension brane should remain
localized within a penetration depth
\begin{equation}
\ell_{RS}^\perp=\frac{3m_4^3}{-2\lambda_3},
\end{equation}
which should be smaller than $0.2\,$mm.
In a reasoning similar to the ADD argument we would
expect an effective 4-dimensional Planck mass
\[
m_3\approx\sqrt{m_4^3\ell_{RS}^\perp},
\]
implying $m_4>10^8\,$GeV, cf.\ (\ref{m4bound}),
and
\begin{equation}\label{lbound}
\lambda_3\approx -\frac{m_3^2}{{\ell_{RS}^\perp}^2}
<-10^{13}\,\mbox{GeV}^4,
\end{equation}
corresponding to a bulk cosmological constant
\begin{equation}\label{Lbound}
\Lambda\approx -\frac{m_3^2}{{\ell_{RS}^\perp}^3}
<-10\,\mbox{GeV}^5.
\end{equation}

It was noticed by M\"uck {\it et al}.\ that a fully fledged
linearized theory indicates that the brane tension should
also be negative in the single brane setup \cite{MVV,AIMVV}.
The negative tension brane is also distinguished by the fact
that timelike geodesics are pulled towards the brane
\cite{MVV,GRS2}:
\[
\frac{d^2x^\perp}{ds^2}=
-\frac{\lambda_3}{6m_4^3}\eta_{\mu\nu}\frac{dx^\mu}{ds}
\frac{dx^\nu}{ds}\mbox{sign}(x^\perp)
\exp\!\left(-\frac{\lambda_3}{3m_4^3}|x^\perp|\right)
\]
\[
\Rightarrow\,\,\mbox{sign}(\ddot{x}^\perp)=
\mbox{sign}(\lambda_3)\mbox{sign}(x^\perp).
\]

If it were for ordinary Einstein--Friedmann cosmology, a
$(3+1)$-dimensional universe with a negative cosmological
constant satisfying (\ref{lbound})
would have collapsed long ago. Here, however, the
effect of $\lambda_3$ is to balance the effect from the bulk
cosmological constant $\Lambda$ to keep the hypersurface
$\mathcal{M}_{3,1}$ flat in the zeroth order approximation.
The fact that $|\lambda_3|\approx |\Lambda|\ell_{RS}^\perp$ 
exceeds effective four-dimensional mean energy densities\footnote{The
parameter $0.6\le h\le 0.8$ parametrizes the uncertainty in
the value of
the Hubble constant $H=100h\,\mbox{km}/(\mbox{s}\,\mbox{Mpc})$.}
\[
\varrho_4\simeq 81h^2\,\mbox{meV}^4,
\]
by at least
a factor $10^{59}$ shows that matter densities can indeed
be treated as perturbations in cosmological
investigations of this scenario.

\subsection{The model of Dvali, Gabadadze and Porrati}\label{dgp}

The idea behind the model of
Dvali, Gabadadze and Porrati is competition between
the bulk curvature scalar $R$
and the corresponding intrinsic curvature scalar 
$R^{(d-1)}$ on the brane \cite{DGP,DG}. We will again focus
on the codimension 1 case and calculate the gravitational
potential.

In the light of the results of Sec.\ \ref{Action} we should write
the action of the model as
\begin{equation}\label{actionDGP}
S=
\frac{m_d^{d-1}}{2}\int dt\int d^{d-1}\vek{x}
\int dx^\perp
\sqrt{-g}R
\end{equation}
\[
+\left.\int dt\int d^{d-1}\vek{x}\,\left(
\frac{m_{d-1}^{d-2}}{2}\sqrt{-g}R^{(d-1)}
-m_d^{d-1}\sqrt{-g}\,
\overline{K}
+\mathcal{L}\right)\right|_{x^\perp=0},
\]
with the Lagrangian $\mathcal{L}$ containing the matter
degrees of freedom. 
The model was motivated by radiative generation
of a kinetic graviton term on the brane
 \cite{DGP,dmc,steve1,steve2}.

The action (\ref{actionDGP}) yields Einstein equations
\[
m_d^{d-1}\left(R_{MN}-\frac{1}{2}g_{MN}R\right)
+m_{d-1}^{d-2}g_M{}^\mu g_N{}^\nu
\left(R^{(d-1)}_{\mu\nu}
-\frac{1}{2}g_{\mu\nu}R^{(d-1)}\right)\delta(x^\perp)
\]
\begin{equation}\label{einsteinDGP}
=
g_M{}^\mu g_N{}^\nu T_{\mu\nu}\delta(x^\perp),
\end{equation}
and the resulting matching condition (\ref{lanczoseq})
is
\begin{equation}\label{lanczosDGP}
\lim_{\epsilon\to +0}
\left[K_{\mu\nu}\right]_{x^\perp =-\epsilon}^{x^\perp =\epsilon}=
\left.\frac{1}{m_d^{d-1}}\left(T_{\mu\nu}-\frac{1}{d-1}
g_{\mu\nu}g^{\alpha\beta}T_{\alpha\beta}\right)\right|_{x^\perp =0}
\end{equation}
\[
-\left.\frac{m_{d-1}^{d-2}}{m_d^{d-1}}\left(R^{(d-1)}_{\mu\nu}
-\frac{1}{2(d-1)}
g_{\mu\nu}g^{\alpha\beta}R^{(d-1)}_{\alpha\beta}
\right)\right|_{x^\perp =0}.
\]
It is amusing that the model of Dvali, Gabadadze and Porrati 
provides a novel and entirely unprecedented
realization of the old proposal of Lorentz and Levi-Civita
to consider $-m_3^2[R^{(3)}_{\mu\nu}-g_{\mu\nu}(R^{(3)}/2)]$
as the energy-momentum tensor of the gravitational field.

 For the weak field approximation I still prefer
to employ Gaussian normal coordinates for the background metric,
because of the inevitable factor $\delta(x^\perp)$ in 
the Einstein equation. This implies that we can impose 
 a harmonic gauge condition
only on the longitudinal coordinates $x^\mu$:
\begin{equation}\label{hggauss}
\partial_\alpha h^\alpha{}_\mu+\partial_\perp h_{\perp\mu}
=\frac{1}{2}\partial_\mu\left(h^\alpha{}_\alpha
+h_{\perp\perp}\right),
\end{equation}
but this is sufficient to get a decoupled equation for
the gravitational potential of a static mass
distribution:\\
The transverse equations in the gauge (\ref{hggauss})
\[
R_{\perp\perp}-R^\alpha{}_\alpha=
\frac{1}{2}\partial_\alpha\partial^\alpha
\left(h^\beta{}_\beta
-h_{\perp\perp}\right)
+\partial_\perp\partial_\alpha h^{\alpha}{}_\perp=0,
\]
\[
R_{\perp\mu}=\frac{1}{2}\left(
\partial_\mu\partial_\alpha h^{\alpha}{}_\perp
-\partial_K\partial^K h_{\perp\mu}\right)
+\frac{1}{4}\partial_\mu\partial_\perp
\left(h_{\perp\perp}-h^\alpha{}_\alpha\right)=0
\]
can be solved by $h_{\perp\mu}=0$, $h_{\perp\perp}=h^\alpha{}_\alpha$,
whence the remaining equations take the form
\[
m_d^{d-1}(\partial_\alpha\partial^\alpha+\partial_\perp^2)h_{\mu\nu}
+m_{d-1}^{d-2}\delta(x^\perp)\left(\partial_\alpha\partial^\alpha
h_{\mu\nu}-\partial_\mu\partial_\nu h^\alpha{}_\alpha\right)
\]
\[
=-2\delta(x^\perp)\left(T_{\mu\nu}
-\frac{1}{d-1}\eta_{\mu\nu}\eta^{\alpha\beta}T_{\alpha\beta}\right).
\]
 For $d=4$ this yields the equation for the gravitational
potential of a 
mass density $\varrho(\vek{r})=M\delta(\vek{r})$ on $\mathcal{M}_{3,1}$:
\begin{equation}\label{UDGP}
m_4^3(\Delta+\partial_\perp^2)U(\vek{r},x^\perp)
+m_3^2\delta(x^\perp)\Delta U(\vek{r},x^\perp)
=\frac{2}{3}M\delta(\vek{r})\delta(x^\perp).
\end{equation}
Insertion of a Fourier {\it ansatz}
\[
U(\vek{r},x^\perp)=\frac{1}{(2\pi)^4}
\int d^3\vek{p}\int dp_\perp\, U(\vek{p},p_\perp) 
\exp\left(i(\vek{p}\cdot\vek{r}+p_\perp x^\perp)\right)
\]
yields an integral equation
\begin{equation}\label{FTU}
m_4^3(\vek{p}^2+p_\perp^2)U(\vek{p},p_\perp)+\frac{m_3^2}{2\pi}
 \vek{p}^2\int dp_\perp'\,U(\vek{p},p_\perp')=-\frac{2}{3}M.
\end{equation}
This equation tells us that $U(\vek{p},p_\perp)$ must be of the form
\[
U(\vek{p},p_\perp)=\frac{f(\vek{p})}{\vek{p}^2+p_\perp^2},
\] 
and $f(\vek{p})$ is then easily determined algebraically:
\begin{equation}\label{FTU2}
U(\vek{p},p_\perp)=-\frac{4}{3}\frac{M}{(\vek{p}^2+p_\perp^2)
(2m_4^3+m_3^2|\vek{p}|)}.
\end{equation}
The resulting potential on the brane is 
\begin{equation}\label{Udgp}
U(\vek{r})=-\frac{M}{6\pi m_3^2 r}\left[
\cos\!\left(\frac{2m_4^3}{m_3^2}r\right)-
\frac{2}{\pi}\cos\!\left(\frac{2m_4^3}{m_3^2}r\right)
\mbox{Si}\!\left(\frac{2m_4^3}{m_3^2}r\right)\right.
\end{equation}
\[
\left.
+\frac{2}{\pi}\sin\!\left(\frac{2m_4^3}{m_3^2}r\right)
\mbox{ci}\!\left(\frac{2m_4^3}{m_3^2}r\right)
\right],
\]
with the sine and cosine integrals
\[
\mbox{Si}(x)=\int_0^xd\xi\,\frac{\sin\xi}{\xi},
\]
\[
\mbox{ci}(x)=-\int_x^\infty d\xi\,\frac{\cos\xi}{\xi}.
\]
The model of Dvali, Gabadadze and Porrati predicts a 
transition scale 
\begin{equation}\label{ldgp}
\ell_{DGP}=\frac{m_3^2}{2m_4^3}
\end{equation}
between four-dimensional behavior 
and five-dimensional behavior
of the gravitational potential:
\begin{eqnarray*}
r\ll\ell_{DGP}:\,\,
U(\vek{r})=\!&-&\!\frac{M}{6\pi m_3^2 r}
\left[1+\left(\gamma-\frac{2}{\pi}\right)\frac{r}{\ell_{DGP}}
\right.\\
&+&\!\frac{r}{\ell_{DGP}}\ln\!\left(\frac{r}{\ell_{DGP}}\right)
+\left.\mathcal{O}\!\left(\frac{r^2}{\ell^2_{DGP}}\right)\right],
\\
r\gg\ell_{DGP}:\,\,
U(\vek{r})=\!&-&\!\frac{M}{6\pi^2 m_4^3 r^2}
\left[1-2\frac{\ell^2_{DGP}}{r^2}
+\mathcal{O}\!\left(\frac{\ell^4_{DGP}}{r^4}\right)\right].
\end{eqnarray*}
$\gamma\simeq 0.577$ is Euler's constant.
If we would use the reduced Planck mass for $m_3$, then
the small $r$ potential would be stronger
than the genuine four-dimensional potential 
by a factor $\frac{4}{3}$ because the coupling of 
the masses on the brane 
to the four-dimensional Ricci tensor is increased
by this factor, cf.\ (\ref{deltaU},\ref{UDGP}).
This factor $\frac{4}{3}$ is in agreement with the
tensorial structure of the graviton propagator reported
in \cite{DGP}, which
has been attributed to an additional helicity state
of the five-dimensional graviton which in a first
approximation appears like mediating an
additional attractive 
scalar interaction from a four-dimensional 
perspective. 
While it might seem like a simple rescaling
of the relation between $m_3$ and $G_{N,3}$, this
additional state is clearly a matter of phenomenological
concern \cite{DGP}.
Note, however, that a
logarithmic modification of the Newton
potential is usually not accounted for in 
the standard parametrized post-Newtonian formalism \cite{will}.
 Furthermore, the logarithmic term  
does not resemble the type of
modification that one expects from an effective
four-dimensional scalar-tensor theory
of gravity. Therefore the phenomenological implications
of the model of Dvali, Gabadadze and Porrati with
 $m_3=(6\pi G_{N,3})^{-1/2}\simeq 2.8\times 10^{18}\,$GeV
warrant further study.

\section{A remark on black holes 
in the model of Dvali, Gabadadze and Porrati}\label{bhi}

Properties of sub-millimeter and primordial black holes in theories with 
sub-millimeter extra dimensions \cite{ADD1,AADD,ADD2}
were discussed by Argyres {\it et al}.\ \cite{ADM}.
Extensions of the Schwarzschild metric
into the bulk of the Randall--Sundrum model
(and variants of it) have been proposed and
investigated in \cite{CHR,GS,DMPR,EHM,GR1,SS,CRSS,naresh,
CW,chamblin,EGS,MPS,NOO}. 

To my knowledge at the time of this writing 
no dedicated investigations of possible
extensions of the Schwarzschild metric or black hole
properties in the framework of the DGP model \cite{DGP}
have been reported.
This may seem surprising given the attractiveness of
this model. However, eqs.\ (\ref{FTU2},\ref{Udgp}) show that even
in the Schwarzschild case, which should translate into an
axially symmetric metric in the DGP model, the result will be
much more complicated than the four- or five-dimensional Schwarzschild
metrics: If an analytic expression can be found at all, it 
inevitably will involve special functions. It is also clear
that the restriction of the metric to the 3-brane will 
approximate the four-dimensional
Schwarzschild metric at most for a certain range of $r$,
where $r$ is supposed to be the standard Schwarzschild radial
coordinate on the brane.

The axial symmetry of (\ref{einsteinDGP}) and spherical symmetry
on the 3-brane imply that on every hypersurface $x^\perp=$ const.\
we should have a radial coordinate $r$ (beyond an eventual
event horizon) such that the sections
$t=$ const., $r=$ const., $x^\perp=$ const.\ correspond to 2-spheres
of circumference $2\pi r$ and area $4\pi r^2$.
This entails a metric {\it ansatz}
\begin{equation}\label{bhansatz}
ds^2=-n^2(x^\perp,r)dt^2+a^2(x^\perp,r)dr^2+r^2d\vartheta^2
+r^2\sin^2\vartheta d\varphi^2+d{x^\perp}^2.
\end{equation}
The coordinates employed in this {\it ansatz} are subject
to the restrictions that $x^\perp$ is only applicable in that
neighborhood of the brane which is covered by geodesics emerging
from the brane, while $r$ must have a lower limit
in terms of an event horizon or the extension of the 
mass distribution generating the metric (\ref{bhansatz}).

With the abbreviations for partial derivatives
\[
\check{f}(x^\perp,r)=\frac{\partial}{\partial r}f(x^\perp,r),
\]
\[
f'(x^\perp,r)=\frac{\partial}{\partial x^\perp}f(x^\perp,r),
\]
the non-vanishing components of the extrinsic curvature tensor
of the hypersurfaces $x^\perp=$ const.\ are
\begin{equation}\label{Ktt}
K_{tt}=nn',
\end{equation}
\begin{equation}\label{Krr}
K_{rr}=-aa',
\end{equation}
and 
we find for the non-vanishing components of the Ricci tensor 
\begin{equation}\label{riccitheta}
R_{\vartheta\vartheta}=
1-\frac{1}{a^2}-\frac{\check{n}r}{na^2}+\frac{\check{a}r}{a^3},
\end{equation}
\begin{equation}\label{ricciphi}
R_{\varphi\varphi}=\sin^2\vartheta R_{\vartheta\vartheta},
\end{equation}
\begin{equation}\label{riccit}
R_{tt}=\frac{n\check{\check{n}}}{a^2}-\frac{n\check{n}\check{a}}{a^3}
+2\frac{n\check{n}}{a^2r}+nn''+\frac{n}{a}n'a'
\end{equation}
\begin{equation}\label{riccir}
R_{rr}=-\frac{\check{\check{n}}}{n}+\frac{\check{n}\check{a}}{na}
+2\frac{\check{a}}{ar}-aa''-\frac{a}{n}n'a',
\end{equation}
\begin{equation}\label{riccirp}
R_{r\perp}=-\frac{\check{n}'}{n}+\frac{a'}{a}\left(\frac{\check{n}}{n}
+\frac{2}{r}\right),
\end{equation}
\begin{equation}\label{riccipp}
R_{\perp\perp}=-\frac{a''}{a}-\frac{n''}{n}.
\end{equation}
The equations (\ref{einsteinDGP},\ref{lanczosDGP}) then translate
into an equation which holds in the whole region of applicability
of the coordinates used
in (\ref{bhansatz}):
\begin{equation}\label{bhall}
\frac{a^2-1}{r}
=\frac{\check{n}}{n}-\frac{\check{a}}{a},
\end{equation}
equations which hold in the bulk:
\begin{equation}\label{bhbulk1}
\frac{a''}{a}=-\frac{n''}{n}
=\frac{1}{a^2r}\left(\frac{\check{a}}{a}
+\frac{\check{n}}{n}\right),
\end{equation}
\begin{equation}\label{bhbulk2}
\frac{\check{\check{n}}}{na^2}-\frac{\check{n}\check{a}}{na^3}
+\frac{n'a'}{na}=\frac{1}{a^2r}\left(\frac{\check{a}}{a}
-\frac{\check{n}}{n}\right),
\end{equation}
\begin{equation}\label{bhbulk3}
\frac{\check{n}'}{n}=\frac{a'}{a}\left(\frac{\check{n}}{n}
+\frac{2}{r}\right),
\end{equation}
and equations holding only on the brane:
\begin{equation}\label{bhb1}
\left[\frac{\check{\check{n}}}{n}-\frac{\check{n}\check{a}}{na}
+\frac{a^2-1}{r^2}\right]_{x^\perp=0}=0,
\end{equation}
\begin{equation}\label{bhb2}
\frac{a'}{a}(x^\perp\to\pm 0,r)=
-\frac{n'}{n}(x^\perp\to\pm 0,r)=
\pm\frac{m_3^2}{2m_4^3}\left[\frac{1}{a^2r}\left(\frac{\check{a}}{a}
+\frac{\check{n}}{n}\right)\right]_{x^\perp=0}.
\end{equation}

These equations allow for a black string solution which would plainly
continue the four-dimensional Schwarzschild metric into the bulk
along the orthogonal geodesics. However, this is an artefact 
of the fact that the coordinates in (\ref{bhansatz}) have an 
event horizon $r_M$,
and it is clearly not the correct solution for a brane black hole:
It would give a four-dimensional Newtonian potential on each
hypersurface $x^\perp=$ const.\ in the large $r$ limit,
instead of fulfilling the
correct boundary condition of a five-dimensional Newtonian potential
at large distance.

 For $r\ll\ell_{DGP}$ we notice that eqs.\ 
(\ref{bhall},\ref{bhb1},\ref{bhb2})
approximate the ordinary equations for the Schwarzschild metric
on the brane, i.e. the correct solution on the brane
will approximate the four-dimensional
Schwarzschild solution for $r_M< r\ll\ell_{DGP}$:
\[
n^2(0,r)\approx 1+2U(r)+\mathcal{O}(r/\ell_{DGP}),
\]
cf.\ (\ref{Udgp},\ref{ldgp}).

 For $r\gg\ell_{DGP}$ we notice that (\ref{bhb2}) implies that
the metric becomes smooth across the brane in that limit,
while (\ref{bhb1}) reduces to a special case of 
(\ref{bhall},\ref{bhbulk2}).
The remaining
equations are just the conditions for a Ricci flat
five-dimensional spacetime, and the solution must approximate an axially
symmetric five-dimensional black hole spacetime, in agreement with
the role of $\ell_{DGP}$ as a scale
 separating four-dimensional effects from five-dimensional effects
in the DGP model.

\section{The cosmology of codimension 1 brane worlds}\label{cosmology}

The five-dimensional Einstein tensor for the line element
(with $x_i\equiv x^i$, $r^2\equiv x_ix^i$)
\begin{equation}\label{cosprinc}
ds^2=-n^2(x^\perp,t)dt^2
\end{equation}
\[
+a^2(x^\perp,t)
\!\left(\delta_{ij}+k\frac{x_i x_j}{1-kr^2}
\right)dx^idx^j
+b^2(x^\perp,t)d{x^\perp}^2
\]
can be found in \cite{BDEL}. Brane cosmology in different backgrounds
or with different {\it ans\"atzen} for the metric has been a subject
of numerous studies. Investigations of cosmology in backgrounds
motivated by $M$-theory or generalizations of
the Randall--Sundrum model can be found in
 \cite{LOW,BDL,kaloper,nihei,CGKT,CGS,CF,FTW,
kraus,lidsey,vollick,CGRT,
neronov,ida,GS,MSM,MWBH,CH0,zurab2,NO,BCG,CKR,CH2}.
Eq.\ (\ref{cosprinc}) implies a brane cosmological
principle in that it presupposes that every hypersurface
 $x^\perp=$ const.\ is a Robertson--Walker spacetime
with cosmological time $T|_{x^\perp}=\int n(x^\perp,t)dt$. 

I will focus on the cosmological aspects
of the model of Dvali, Gabadadze and Porrati.
Building on the results of \cite{BDL,BDEL},
the evolution equations of a 3-brane in 
a five-dimensional bulk following from
(\ref{einsteinDGP}) and (\ref{lanczosDGP}) were so neatly presented in
recent papers by Deffayet \cite{cedric} and by Deffayet, Dvali
and Gabadadze \cite{DDG} that I decided to give the corresponding results
for a $\nu$-brane\footnote{This moderate generalization spares me from the
frustrating experience of plainly repeating the equations of Deffayet
{\it et al}. It also may be of some interest in its own to
have the corresponding equations for a $(\nu+1)$-dimensional
timelike hypersurface at hand.}.
 
 The Einstein tensors for the metric (\ref{cosprinc})
in Gaussian normal coordinates ($b^2=1$) and in $d=\nu+1$ 
spatial dimensions are\\
on the hypersurfaces $x^\perp=$ const.:
\begin{equation}\label{gbrane00}
G_{00}^{(\nu)}=\frac{1}{2}\nu(\nu-1)n^2
\!\left(\frac{\dot{a}^2}{n^2a^2}+\frac{k}{a^2}\right)
\end{equation}
\begin{eqnarray}\label{gbraneij}
G_{ij}^{(\nu)}=(\nu-1)\!\left(
\frac{\dot{n}\dot{a}}{n^3a}-\frac{\ddot{a}}{n^2a}\right)g_{ij}
-\frac{1}{2}(\nu-1)(\nu-2)\!\left(
\frac{\dot{a}^2}{n^2a^2}+\frac{k}{a^2}\right)g_{ij},
\end{eqnarray}
and in the bulk:
\begin{equation}\label{bdl00}
G_{00}=\frac{1}{2}\nu(\nu-1)n^2
\!\left(\frac{\dot{a}^2}{n^2a^2}-\frac{{a'}^2}{a^2}
+\frac{k}{a^2}\right)-\nu n^2\frac{a''}{a},
\end{equation}
\begin{eqnarray}\label{bdlij}
G_{ij}&=&\frac{1}{2}(\nu-1)(\nu-2)\!\left(\frac{{a'}^2}{a^2}
-\frac{\dot{a}^2}{n^2a^2}-\frac{k}{a^2}\right)g_{ij}
\\
 &&\!\! +(\nu-1)\!\left(\frac{a''}{a}+\frac{n'a'}{na}
-\frac{\ddot{a}}{n^2a}+\frac{\dot{n}\dot{a}}{n^3a}\right)g_{ij}
+\frac{n''}{n}g_{ij},\nonumber
\end{eqnarray}
\begin{equation}\label{bdl0perp}
G_{0\perp}=\nu\!\left(
\frac{n'}{n}\frac{\dot{a}}{a}-\frac{\dot{a}'}{a}
\right),
\end{equation}
\begin{equation}\label{bdlpp}
G_{\perp\perp}=\frac{1}{2}\nu(\nu-1)
\!\left(\frac{{a'}^2}{a^2}-\frac{\dot{a}^2}{n^2a^2}
-\frac{k}{a^2}\right)
+\nu\!\left(\frac{n'a'}{na}+\frac{\dot{n}\dot{a}}{n^3a}
-\frac{\ddot{a}}{n^2a}\right).
\end{equation}

The matching conditions (\ref{lanczosDGP}) for
an ideal fluid on the brane 
\[
T_{00}=\varrho n^2,\,\, T_{ij}=p g_{ij}
\]
read
\begin{eqnarray}\label{coslanczoseq00}
\lim_{\epsilon\to +0}\left[
\partial_\perp n\right]_{x^\perp =-\epsilon}^{x^\perp =\epsilon}
&=&\frac{n}{\nu m_{\nu+1}^\nu}\left.\!\Bigg(
(\nu-1)\varrho+\nu p\Bigg)\right|_{x^\perp =0}\\
&&\!\! +\frac{m_{\nu}^{\nu-1}}{m_{\nu+1}^\nu}(\nu-1)n\left.\!\left(
\frac{\ddot{a}}{n^2a}-\frac{\dot{a}^2}{2n^2a^2}-
\frac{\dot{n}\dot{a}}{n^3a}-\frac{k}{2a^2}\right)\right|_{x^\perp =0},
\nonumber
\end{eqnarray}
\begin{equation}\label{coslanczoseqij}
\lim_{\epsilon\to +0}\left[
\partial_\perp a\right]_{x^\perp =-\epsilon}^{x^\perp =\epsilon}
=\frac{m_{\nu}^{\nu-1}}{2m_{\nu+1}^\nu}(\nu-1)\left.\!\left(
\frac{\dot{a}^2}{n^2a}+\frac{k}{a}\right)\right|_{x^\perp =0}
-\left.\frac{\varrho a}{\nu m_{\nu+1}^\nu}\right|_{x^\perp =0}.
\end{equation}
In the spirit of the remark following eq.\ (\ref{lanczosDGP}) this
corresponds to effective gravitational contributions to
the pressure and energy density on the brane:
\[
\varrho_G=-\frac{1}{2}\nu(\nu-1)m_{\nu}^{\nu-1}\!\left(
\frac{\dot{a}^2}{n^2a^2}+\frac{k}{a^2}\right),
\]
\[
p_G=(\nu-1)m_{\nu}^{\nu-1}\!\left(
\frac{\ddot{a}}{n^2a}-\frac{\dot{n}\dot{a}}{n^3a}\right)
+\frac{1}{2}(\nu-1)(\nu-2)m_{\nu}^{\nu-1}\!\left(
\frac{\dot{a}^2}{n^2a^2}+\frac{k}{a^2}\right).
\]

Not surprisingly, energy conservation on the brane follows from
the absence of transverse momentum, $T_{0\perp}=0$.
With (\ref{bdl0perp}) this implies
\begin{equation}\label{0perp0}
\frac{n'}{n}=\frac{\dot{a}'}{\dot{a}}
\end{equation}
and in particular
\[
\lim_{\epsilon\to +0}\left[\frac{n'}{n}
\right]_{x^\perp =-\epsilon}^{x^\perp =\epsilon}
=\lim_{\epsilon\to +0}\left[\frac{\dot{a}'}{\dot{a}}
\right]_{x^\perp =-\epsilon}^{x^\perp =\epsilon}.
\]
Insertion of (\ref{coslanczoseq00},\ref{coslanczoseqij})
into this equation yields the sought for conservation
equation
\begin{equation}\label{econ}
\dot{\varrho}a\Big|_{x^\perp=0}=-\nu(\varrho+p)\dot{a}\Big|_{x^\perp=0}.
\end{equation}

Insertion of (\ref{0perp0}) into (\ref{bdl00}) and (\ref{bdlpp})
for $x^\perp\neq 0$
yields a $\nu$-dimensional version of the integral
of Bin\'{e}truy {\it et al}.\ \cite{BDEL}:
\[
\frac{2}{\nu n^2}a'a^\nu G_{00}=
\frac{\partial}{\partial x^\perp}\!\left(
\frac{\dot{a}^2}{n^2}a^{\nu-1}-
{a'}^2a^{\nu-1}+ka^{\nu-1}
\right)=0,
\]
\[
\frac{2}{\nu}\dot{a}a^\nu G_{\perp\perp}=
-\frac{\partial}{\partial t}\!\left(
\frac{\dot{a}^2}{n^2}a^{\nu-1}-
{a'}^2a^{\nu-1}+ka^{\nu-1}
\right)=0,
\]
i.e.
\begin{equation}\label{intbdel+}
I^{+}_{BDEL}=\left.
\left(\frac{\dot{a}^2}{n^2}-
{a'}^2+k\right)a^{\nu-1}\right|_{x^\perp >0}
\end{equation}
and
\begin{equation}\label{intbdel-}
I^{-}_{BDEL}=\left.
\left(\frac{\dot{a}^2}{n^2}-
{a'}^2+k\right)a^{\nu-1}\right|_{x^\perp <0}
\end{equation}
are two constants, with $I^{+}_{BDEL}=I^{-}_{BDEL}$
if 
\[
\lim_{\epsilon\to +0}a'\Big|_{x^\perp=\epsilon}
=\pm\lim_{\epsilon\to +0}a'\Big|_{x^\perp=-\epsilon}.
\]

We have not yet taken into account $G_{ij}=0$ in the bulk.
However, eq.\ (\ref{0perp0}) implies 
$\partial_\perp(n/\dot{a})=0$, and therefore
\[
\frac{n''}{n}=\frac{\dot{a}''}{\dot{a}}.
\]
This, the bulk equations $G_{00}=G_{\perp\perp}=0$, and the constancy
of $I^{\pm}$ imply that the bulk equation $G_{ij}=0$
is already satisfied and does not provide any new information.

We can now simplify the previous equations by further restricting
our Gaussian normal coordinates through the gauge
\begin{equation}\label{frwgauge}
n(0,t)=1
\end{equation}
by simply performing the transformation
\[
t\,\,\Rightarrow\,\,t_{FRW}=\int^tdt'\,n(0,t')
\]
of the time coordinate.
This gauge is convenient because it gives the usual
cosmological time on the brane. Henceforth this gauge will be
adopted, but the index FRW will be omitted.

Eqs.\ (\ref{0perp0},\ref{frwgauge}) imply that our 
set of dynamical variables
is not $\{n(x^\perp,t),a(x^\perp,t)\}$ but
only $a(x^\perp,t)$, with $n(x^\perp,t)$ given by
\[
n(x^\perp,t)=\frac{\dot{a}(x^\perp,t)}{\dot{a}(0,t)}.
\]
The basic set of cosmological equations in the present setting 
(without a cosmological
constant in the bulk) are thus
eqs.\ (\ref{coslanczoseqij},\ref{econ},\ref{intbdel+},\ref{intbdel-}),
 which have to be amended
with dispersion relations (or corresponding evolution equations) for the 
ideal fluid
components on the brane:\\[2ex]
\fbox{
\begin{minipage}{0.97\textwidth}
\vspace*{2ex}
\[
\lim_{\epsilon\to +0}\left[
\partial_\perp a\right]_{x^\perp =-\epsilon}^{x^\perp =\epsilon}(t)
=\frac{m_{\nu}^{\nu-1}}{2m_{\nu+1}^\nu}(\nu-1)
\frac{\dot{a}^2(0,t)+k}{a(0,t)}
-\frac{\varrho(t) a(0,t)}{\nu m_{\nu+1}^\nu},
\]
\[
I^{\pm}_{BDEL}=\left.
\left(\dot{a}^2(0,t)-
{a'}^2(x^\perp,t)+k\right)a^{\nu-1}(x^\perp,t)\right|_{x^\perp\gtrless 0},
\]
\[
\dot{\varrho}(t)a(0,t)=-\nu(\varrho(t)+p(t))\dot{a}(0,t),
\]
\[
p(t)=p(\varrho(t)).
\]
\vspace*{1ex}
\end{minipage}
}

\vspace*{2ex}
Our primary concern with regard to observational consequences is the
evolution of the scale factor $a(0,t)$ on the brane, and we can use
{\it les int\'{e}grales fran\c{c}aises} $I^{\pm}$ to eliminate
the normal derivatives $a'(x^\perp\to\pm 0,t)$ from the brane analogue 
of the Friedmann equation:
\[
\pm\sqrt{\dot{a}^2(0,t)+k-I^+_{BDEL}a^{1-\nu}(0,t)}
\mp\sqrt{\dot{a}^2(0,t)+k-I^-_{BDEL}a^{1-\nu}(0,t)}
\]
\begin{equation}\label{friedmann1}
=\frac{m_{\nu}^{\nu-1}}{2m_{\nu+1}^\nu}(\nu-1)
\frac{\dot{a}^2(0,t)+k}{a(0,t)}
-\frac{\varrho(t) a(0,t)}{\nu m_{\nu+1}^\nu}.
\end{equation}
If this equation is solved for $a(0,t)$
by using the dispersion relation and energy
conservation on the brane, then $a(x^\perp,t)$ can be determined in the
bulk from the constancy of $I^{\pm}$.

There must be at least one minus sign on the left hand side of 
(\ref{friedmann1}) if the right hand side is negative, but the dynamics
of the problem does not require symmetry across the brane.
The constants $I^{\pm}$ must be considered as initial conditions,
and if e.g.\
 $I^+\neq I^-$, then there cannot be any symmetry across the brane.

If $m_\nu\neq 0$ and the 
normal derivatives on the
brane have the same sign:
\begin{equation}\label{ordincond}
m_\nu a'(x^\perp\to +0,t)a'(x^\perp\to -0,t)>0,
\end{equation}
 then the cosmology of our brane approximates
ordinary Friedmann--Robertson--Walker cosmology during those epochs
when
\[
I^{\pm}_{BDEL}\ll \left(\dot{a}^2(0,t)+k\right)a^{\nu-1}(0,t).
\]
In particular, this applies to late epochs in expanding open or flat
branes ($k\neq 1$).

However, standard cosmology may be realized in this model
in an even more direct way:
If (\ref{ordincond}) holds
and $I^+=I^-$, then (\ref{friedmann1}) reduces {\em entirely} to 
the ordinary Friedmann equation for a $(\nu+1)$-dimensional
spacetime. The evolution of the background geometry of the observable
universe according to the Friedmann equation can thus be embedded
 in the model of Dvali, Gabadadze and Porrati, with the bevavior 
of $a(x^\perp,t)$ off the brane determined solely by the integral 
 $I^+=I^-$ and the boundary condition $a(0,t)$ from the Friedmann
equation. 

This possibility of a direct embedding of Friedmann cosmology 
is a consequence of the fact that the evolution of the background
geometry (\ref{cosprinc}) and the source terms $\varrho$, $p$
are supposed to depend only on $t$ and $x^\perp$.
This implies the possibility to decouple the brane and the bulk contributions
in the Einstein equation for the background metric, and
in this case deviations
from Friedmann--Robertson--Walker cosmology would only show up
in specific $\vek{x}$-dependent effects like the evolution of
cosmological perturbations and structure formation.

As a simple example of the realization of this direct embedding
of a Friedmann--Robertson--Walker 3-brane in the model of Dvali
{\it et al}.\ we consider a spatially flat ($k=0$) radiation dominated
 ($p=\varrho/3$) 3-brane with continuous normal derivative
$a'$ across the brane:

Since $a$ is smooth across the brane we have $I^+=I^-$ and the
signs in (\ref{friedmann1}) conspire in such a way that the left hand side
vanishes. With $\nu=3$ the right hand side boils down to the
ordinary Friedmann equation in a spatially flat radiation dominated
 background, with solution
\[
\varrho(t)=\frac{3m_3^2}{4t^2},
\]
\begin{equation}\label{initcond}
a(0,t)=C\sqrt{t}.
\end{equation}
The integral of Bin\'{e}truy {\it et al}.\ then yields the differential
equation
\[
a^2(x^\perp,t){a'}^2(x^\perp,t)+I_{BDEL}=\frac{C^2}{4t}a^2(x^\perp,t),
\]
which has to be solved under the boundary condition (\ref{initcond}).
This yields
\[
a^2(x^\perp,t)=\frac{C^2}{4t}{x^\perp}^2+\sqrt{C^4-4I_{BDEL}}x^\perp
+C^2t
\]
and
\[
n^2(x^\perp,t)=\frac{C^2}{4t^2}
\frac{\left(4t^2-{x^\perp}^2\right)^2}{C^2{x^\perp}^2
+4\sqrt{C^4-4I_{BDEL}}x^\perp t+4C^2t^2}.
\]
There is a coordinate singularity on the spacelike hypersurfaces
 $x^\perp=\pm 2t$, which indicates that the orthogonal geodesics emerging
from the 3-brane do not cover the full five-dimensional manifold,
but our Gaussian normal coordinates were anyhow expected to cover
only a neighborhood of the brane.

The possibility to describe the cosmological evolution of the
3-brane background geometry by an ordinary Friedmann equation
implies that we will have to rely on specifically $\vek{x}$-dependent
effects to observationally distinguish Friedmann cosmology
from brane cosmology.



\end{document}